\documentclass[aps,prl,superscriptaddress,showpacs,twocolumn]{revtex4}
\usepackage{graphics}
\usepackage{amsmath}
\usepackage{amsfonts}
\usepackage{graphicx}

\begin{document}

\title{Macroscopic entanglement by entanglement swapping}

\author{Stefano Pirandola}
\author{David Vitali}
\author{Paolo Tombesi}
\affiliation{Dipartimento di Fisica, Universit\`{a} di Camerino,
I-62032 Camerino, Italy.}
\author{Seth Lloyd}
\affiliation{Research Laboratory of Electronics and Department of
Mechanical Engineering, Massachusetts Institute of Technology, MA
02139, USA.}

\date{\today}

\begin{abstract}
We present a scheme for entangling two micromechanical
oscillators. The scheme exploits the quantum effects of radiation
pressure and it is based on a novel application of entanglement
swapping, where standard optical measurements are used to generate
purely mechanical entanglement. The scheme is presented by first
solving the general problem of entanglement swapping between
arbitrary bipartite Gaussian states, for which simple input-output
formulas are provided.
\end{abstract}


\maketitle

Micromechanical resonators with fundamental vibrational mode
frequencies in the range 10 MHz--1 GHz can now be fabricated
\cite{Roukes03}. Applications include fast, ultrasensitive
displacement detectors \cite{cleland03}, electrometers, and radio
frequency signal processors. Advances in the development of
micromechanical devices also raise the fundamental question of
whether mechanical systems containing macroscopic numbers of atoms
will exhibit quantum behavior. In particular, it is interesting to
see under which conditions it is feasible to prepare
micromechanical oscillators in \emph{entangled} states, where the
quantum nature becomes most manifest. Proposals of this kind
recently appeared in the literature. One could entangle a
nanomechanical oscillator with a Cooper-pair box \cite{Armour03},
with an ion \cite{zoller}, or with single photons
\cite{Marshall03}; moreover Ref.~\cite{eisert} studied how to
entangle an array of nanomechanical oscillators,
Ref.~\cite{CRITERIO} proposed to entangle two mirrors of an
optical ring cavity, while Ref.~\cite {Peng03} considered two
mirrors of two different cavities illuminated with entangled light
beams. In this Letter we will show how to entangle two
micromechanical oscillators, which are parts of two different
micro-opto-mechanical systems, by means of continuous variable
(CV) entanglement swapping performed on the ``optical parts''. Our
proposal exploits some results of Ref.~\cite{PRLePRA}, which
showed how radiation pressure can entangle a vibrational mode of a
mirror with the back-scattered sideband modes of an intense
optical beam. Thanks to this optomechanical entanglement, one can
entangle two different mechanical oscillators by performing
appropriate optical measurements. Note that, very recently,
Ref.~\cite{Carmon05} has experimentally observed the two
considered sideband modes, although no measurements of
entanglement have been performed. The proposal is presented by
first solving the general problem of CV entanglement swapping
between arbitrary Gaussian bipartite states, for which we are able
to derive compact input-output formulas. Such a study suitably
generalizes the previous \emph{fully optical} theoretical analyses
\cite{vanloock} and it has a straightforward application to our
physical scheme.

Let us first study entanglement swapping with Gaussian states in general. We
consider a pair of CV systems labelled by $k=1,2$, where each of them is
described by a pair of conjugate dimensionless quadratures $\hat{x}_{k}$ and
$\hat{p}_{k}$. Introducing the vector $\hat{\xi}^{T}\equiv (\hat{x}_{1},\hat{%
p}_{1},\hat{x}_{2},\hat{p}_{2})$, we can write the canonical commutation
relations in the compact form $[\hat{\xi}_{l},\hat{\xi}_{m}]=i\mathcal{J}%
_{lm}$ ($l,m=1,...,4$), where
\begin{equation}
\mathcal{J}\equiv J\oplus J,~J\equiv \left(
\begin{array}{cc}
0 & 1 \\
-1 & 0
\end{array}
\right) \text{ ,}  \label{sympl2}
\end{equation}
and $\oplus $ denotes the usual direct sum operation. An arbitrary state $%
\rho _{12}$ of the two oscillators can be described by its Wigner
characteristic function $\Phi _{12}(\vec{\zeta})\equiv
\mathrm{Tr}[\rho _{12}\hat{D}(\vec{\zeta})]$, where
$\hat{D}(\vec{\zeta})=\exp (-i\hat{\xi}^{T} \vec{\zeta})$ is the
Weyl operator and $\vec{\zeta}\in \mathbb{R}^{4}$. Such a state is
said to be Gaussian if the corresponding characteristic function
is Gaussian, i.e., $\Phi _{12}(\vec{\zeta })=\exp
(-\vec{\zeta}^{T}V\vec{\zeta}/2+i\vec{d}^{T}\vec{\zeta})$. In such
a case,
the state $\rho _{12}$ is fully characterized by its displacement $\vec{d}%
\equiv \langle \hat{\xi}\rangle $ and its correlation matrix (CM) $V$, whose
generic element is defined as $V_{lm}\equiv \langle \Delta \hat{\xi}%
_{l}\Delta \hat{\xi}_{m}+\Delta \hat{\xi}_{m}\Delta \hat{\xi}_{l}\rangle /2$
where $\Delta \hat{\xi}_{l}\equiv \hat{\xi}_{l}-\langle \hat{\xi}_{l}\rangle
$. All the information about the quantum correlations between the two
oscillators is contained in the CM, which is a $4\times 4$, real and
symmetric matrix, satisfying the uncertainty principle $V+i\mathcal{J}/2\geq
0$. Putting the CM into the form
\begin{equation}
V\equiv \left(
\begin{array}{cc}
A & D \\
D^{T} & C
\end{array}
\right) \text{ ,}  \label{blocks}
\end{equation}
where $A,C,$ and $D$ are $2\times 2$ real matrices, one can easily derive a
quantitative measure of entanglement. In fact, from the CM~(\ref{blocks}),
one can compute the two quantities
\begin{equation}
\eta ^{\pm }\equiv 2^{-1/2}\left[ \Sigma (V)\pm \left( \Sigma (V)^{2}-4\det
V\right) ^{1/2}\right] ^{1/2}\text{ ,}  \label{Sympl_eigenv}
\end{equation}
where $\Sigma (V)\equiv \det A+\det C-2\det D$. Such quantities correspond
to the \emph{symplectic eigenvalues} \cite{Salerno1} of the matrix $\Lambda
_{b}V\Lambda _{b}$, which is obtained from $V$\ through the partial
transposition (PT) transformation $\Lambda _{b}\equiv \mathrm{\Delta }%
[1,1,1,-1]$, where $\mathrm{\Delta }[\lambda _{1},...,\lambda _{n}]$ denotes
a diagonal matrix with the $n$ entries $\lambda _{k}$ in the diagonal. One
can then prove \cite{Salerno1} that the minimum PT symplectic eigenvalue $%
\eta ^{-}$\ represents an entanglement monotone, since it is connected to
the logarithmic negativity by
\begin{equation}
E_{\mathcal{N}}=\max [0,-\ln 2\eta ^{-}]\text{ .}  \label{logneg}
\end{equation}
In particular, the Gaussian state $\rho _{12}$ is entangled if and only if $%
\eta ^{-}<1/2$, which is equivalent to $E_{\mathcal{N}}>0$ according to Eq.~(%
\ref{logneg}).

In general, CV entanglement swapping involves two pairs of modes, $(a,c_{1})$%
, owned by Alice and Charlie, and $(b,c_{2})$, owned by Bob and Charlie. We
assume that the two pairs are described by two arbitrary Gaussian states $%
\rho _{ac}$ and $\rho _{bc}$, having generic CMs
\begin{equation}
V_{ac}=\left(
\begin{array}{cc}
A & D_{1} \\
D_{1}^{T} & C_{1}
\end{array}
\right) ,\text{ }V_{bc}=\left(
\begin{array}{cc}
B & D_{2} \\
D_{2}^{T} & C_{2}
\end{array}
\right) \text{ ,}  \label{CMs}
\end{equation}
where $A,B,C_{1},C_{2},D_{1},D_{2}$ are $2\times 2$ real matrices. The basic
idea of entanglement swapping is to transfer the bipartite entanglement
within the \emph{near} pairs Alice-Charlie and Charlie-Bob to the\emph{\
distant} pair Alice-Bob, by means of a suitable local operation and
classical communication performed by Charlie. In fact, Charlie mixes his two
modes $c_{1}$ and $c_{2}$ through a balanced beam splitter (BS) and then he
measures the output quadratures $\hat{x}_{c_{2}}-\hat{x}_{c_{1}}$ and $\hat{p%
}_{c_{2}}+\hat{p}_{c_{1}}$. After this measurement, which realizes a CV
version of a Bell measurement, he communicates the outcomes to Alice and Bob
\cite{vanloock}. The final output state $\rho _{ab}$ of Alice and Bob turns
out to be a Gaussian state whose CM can be expressed in terms of the input
matrices' blocks as
\begin{align}
V_{ab}& =V_{ab}^{tr}-g\left(
\begin{array}{cc}
D_{1} &  \\
& D_{2}
\end{array}
\right) \mathcal{J}\left[ \Lambda _{b}\left(
\begin{array}{cc}
C_{1} & C_{1} \\
C_{1} & C_{1}
\end{array}
\right) \Lambda _{b}\right.  \notag \\
& \left. ~~~~~~~~+\Lambda _{a}\left(
\begin{array}{cc}
C_{2} & C_{2} \\
C_{2} & C_{2}
\end{array}
\right) \Lambda _{a}\right] \mathcal{J}^{T}\left(
\begin{array}{cc}
D_{1}^{T} &  \\
& D_{2}^{T}
\end{array}
\right) \text{ ,}  \label{Vout}
\end{align}
where $V_{ab}^{tr}\equiv A\oplus B$, $\Lambda _{a}\equiv R\oplus I$ and $%
\Lambda _{b}\equiv I\oplus R$ are the PT transformations over Alice and Bob
respectively, with $R=\mathrm{\Delta }[1,-1]$ and $I=\mathrm{\Delta }[1,1]$,
and finally $g^{-1}=\det C_{1}+\det C_{2}+\mathrm{Tr}(RC_{2}RJC_{1}J^{T})$.
The general result of Eq.~(\ref{Vout}) specializes to a very simple form if
the input matrices $V_{ab}$ and $V_{bc}$ are put in the normal form, i.e., $%
A=aI,$ $B=bI,$ $C_{k}=c_{k}I$ and $D_{k}=\mathrm{\Delta }[d_{k},d_{k}^{%
\prime }]$ for $k=1,2$. This can always be achieved by means of \emph{local}
symplectic transformations \cite{DuanPRL} at Alice, Bob and Charlie's sites,
which therefore do not affect the entanglement resources of the initial
states $\rho _{ac}$ and $\rho _{bc}$. In such a case, we have
\begin{eqnarray}
V_{ab} &=&\left(
\begin{array}{cc}
aI &  \\
& bI
\end{array}
\right) -\frac{1}{c_{1}+c_{2}}\times  \label{Vab_normalForm} \\
&&\left(
\begin{array}{cc}
\mathrm{\Delta }[d_{1}^{2},d_{1}^{\prime 2}] & \mathrm{\Delta }%
[-d_{1}d_{2},d_{1}^{\prime }d_{2}^{\prime }] \\
\mathrm{\Delta }[-d_{1}d_{2},d_{1}^{\prime }d_{2}^{\prime }] & \mathrm{%
\Delta }[d_{2}^{2},d_{2}^{\prime 2}]
\end{array}
\right) \text{ ,}  \notag
\end{eqnarray}
providing a straightforward expression for the CM of the final Alice and
Bob's state.

In general, the output entanglement \emph{cannot} be expressed as
a simple function of the input entanglement. Nonetheless, this
becomes possible in the symmetrical case of identical physical
resources shared by the two pairs, i.e., when $\rho _{ac}=\rho
_{bc}\equiv \rho _{in}$. If the input matrix $V_{in}$\ of $\rho
_{in}$ is converted from the general form of Eq.~(\ref{blocks}) to
its normal form, the transformation connecting $V_{in}$ to the
output matrix $V_{out}$ of $\rho _{out}\equiv \rho _{ab}$ can be
written as
\begin{eqnarray}
V_{in} &\equiv &\left(
\begin{array}{cc}
aI & \mathrm{\Delta }[d,d^{\prime }] \\
\mathrm{\Delta }[d,d^{\prime }] & cI
\end{array}
\right) \longrightarrow V_{out}=\frac{1}{2c}\times  \label{Transf} \\
&&\left(
\begin{array}{cc}
\mathrm{\Delta }[2ac-d^{2},2ac-d^{\prime 2}] & \mathrm{\Delta }%
[d^{2},-d^{^{\prime }2}] \\
\mathrm{\Delta }[d^{2},-d^{^{\prime }2}] & \mathrm{\Delta }%
[2ac-d^{2},2ac-d^{\prime 2}]
\end{array}
\right) \text{.}  \notag
\end{eqnarray}
Note that in Eq.~(\ref{Transf}) the output state is a symmetric
Gaussian state, whose entanglement properties have been widely
studied in literature \cite{Cirac}. By applying
Eq.~(\ref{Sympl_eigenv}) to $V_{in}$ and $V_{out}$, one can
connect the PT symplectic eigenvalues $\eta _{out}^{\pm }$ of the
output matrix with the ones $\eta _{in}^{\pm }$ of the input
matrix by the simple \emph{input-output}\ relations
\begin{align}
\eta _{out}^{-}& =c^{-1}\eta _{in}^{+}\eta _{in}^{-}=c^{-1}\sqrt{\det V_{in}}%
\text{ ,}  \label{etam} \\
\eta _{out}^{+}& =a\text{ .}  \label{etap}
\end{align}
Note that the output eigenvalues are uniquely determined by the local
symplectic \emph{invariants} of the original CM of Eq.~(\ref{blocks}), since
here $a=\sqrt{\det A}$ and $c=\sqrt{\det C}$.

\begin{figure}[tbp]
\begin{center}
\includegraphics[width=0.37\textwidth]{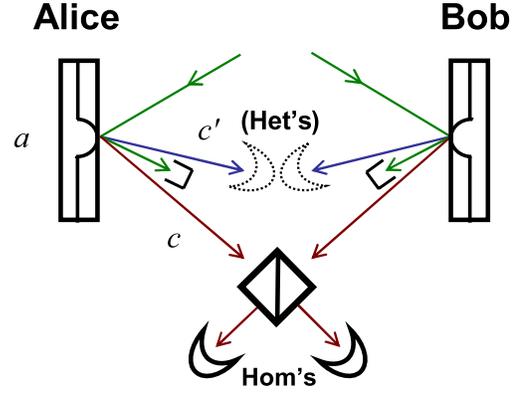}
\end{center}
\par
\vspace{-0.1cm} \caption{Two identical intense laser pulses are
incident on two identical micromechanical oscillators. Each beam
excites a vibrational mode, which generates two optical sideband
modes in the back-scattered field. Both reflected Stokes modes are
mixed through a balanced BS and the output ports are homodyned.
The anti-Stokes modes can be either ignored (non-assisted
strategy) or heterodyned (assisted strategy).} \label{setup1color}
\end{figure}

Let us now apply the above results and, in particular, the input-output
relation of Eq.~(\ref{etam}), to see how one can entangle two
micromechanical oscillators by means of entanglement swapping. Consider the
system of Ref.~\cite{PRLePRA}, which is schematically depicted in the left
part of Fig.~\ref{setup1color}. The radiation pressure of an intense
monochromatic laser beam (frequency $\omega $), incident on a
micromechanical oscillator, generates an effective coupling between a
vibrational mode of the oscillator (label $a$, frequency $\Omega $) and the
two first optical sideband modes induced by the vibrations in the
back-scattered field, i.e., the Stokes mode (label $c$, frequency $\omega
-\Omega $) and the anti-Stokes mode (label $c^{\prime }$, frequency $\omega
+\Omega $) (e.g., see Ref.~\cite{Carmon05}, where these sidebands have been
generated and detected). The vibrational mode is characterized by a
relaxation time $\Gamma ^{-1}$ which globally takes into account several
damping effects, like internal losses of the medium and those due to the
clamping. All these damping effects can be neglected if the duration of the
laser pulse is much shorter than $\Gamma ^{-1}$\ so that the system dynamics
can be assumed unitary during the interaction time. In this unitary
description of the process, the interaction Hamiltonian for the vibrational
mode, the Stokes mode and the anti-Stokes mode is given by ${\hat{H}}%
=-i\hbar \chi (\hat{a}\hat{c}-\hat{a}^{\dag }\hat{c}^{\dag })-i\hbar \theta (%
\hat{a}^{\dag }\hat{c}^{\prime }-\hat{a}\hat{c}^{\prime \dag })$, where $%
\chi $ and $\theta $ are coupling constants whose ratio $r\equiv \theta
/\chi =\left[ (\omega +\Omega )/(\omega -\Omega )\right] ^{1/2}\geq 1$
depends only on the involved frequencies, and the operators $\hat{a}$, $%
\hat{c}$ and $\hat{c}^{\prime }$ are the annihilation operators of the three
modes $a$, $c$ and $c^{\prime }$, satisfying $\left[ \hat{a},\hat{a}%
^{\dagger }\right] =\left[ \hat{c},\hat{c}^{\dagger }\right] =1$, etcetera.
Starting from vacuum states for the optical modes and a thermal state for
the vibrational mode, with mean excitation number $\bar{n}=[\exp (\hbar
\Omega /KT)-1]^{-1}$ ($T$ being the temperature), the evolved state is
Gaussian with the $6\times 6$ CM
\begin{equation}
V_{cac^{\prime }}=\left(
\begin{array}{ccc}
(\mathcal{A}+1/2)I & \mathcal{C}R & \mathcal{F}R \\
\mathcal{C}R & (\mathcal{B}+1/2)I & -\mathcal{D}I \\
\mathcal{F}R & -\mathcal{D}I & (\mathcal{E}+1/2)I
\end{array}
\right) \text{ ,}  \label{Vacc_prime}
\end{equation}
where $\mathcal{A}$, $\mathcal{B}$, $\mathcal{C}$, $%
\mathcal{D}$, $\mathcal{E}$ and $\mathcal{F}$, explicitly given in Ref.~\cite{PRLePRA},
depend on $r$, $\bar{n}$ and the scaled time $t^{\prime }\equiv
t(\theta ^{2}-\chi ^{2})^{1/2}$.

\begin{figure}[tbp]
\begin{center}
\includegraphics[width=0.49\textwidth]{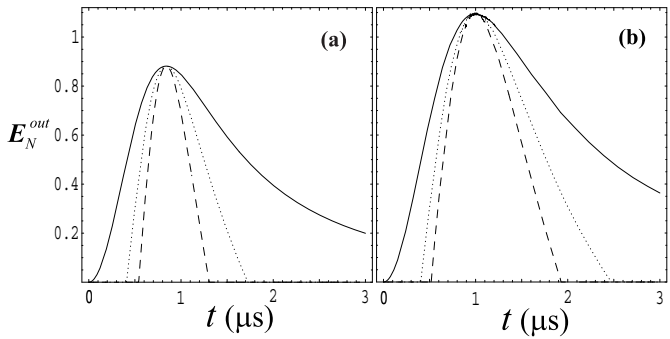}
\end{center}
\par
\vspace{-0.2cm}
\caption{Output log-negativity $E_{\mathcal{N}}^{out}$ versus time $t$, for $%
T=0$K (solid), $T=5$mK (dotted), $T=300$K (dashed) (see text for the other
parameters). (a) refers to the non-assisted strategy; (b) refers to the
assisted strategy.}
\label{Plot}
\end{figure}

Suppose that Alice and Bob are supplied with two identical micromechanical
oscillators, equally illuminated by an intense laser pulse, and suppose that
the reflected Stokes and anti-Stokes modes are sent to Charlie (i.e., to our
experimental detection apparatus). In a first strategy (see Fig.~\ref
{setup1color}), Charlie simply traces out the anti-Stokes modes and performs
a CV Bell measurement over the remaining Stokes modes. In such a case, just
after reading out the measurement results at time $t$, the two mechanical
oscillators are described by a Gaussian state $\rho _{out}$ with CM $V_{out}$
given in Eq.~(\ref{Transf}) through the replacements: $a=\mathcal{B}+1/2,$ $%
c=\mathcal{A}+1/2$ and $d=-d^{\prime }=\mathcal{C}$. By applying Eq.~(\ref
{etam}), one can compute the entanglement $\eta _{out}^{-}$ of the output
state. Adopting the parameters $(\theta ^{2}-\chi ^{2})^{1/2}\simeq 10^{3}$ s%
$^{-1}$, $r=1+2.5\times 10^{-7}$ and $\Omega =5\times 10^{8}$ s$^{-1}$, we
have studied the output log-negativity $E_{\mathcal{N}}^{out}=\max [0,-\ln
2\eta _{out}^{-}]$ as a function of the interaction time $t$, and for
different values of the external temperature $T$. In Fig.~\ref{Plot}(a) one
can see the existence of wide non-zero time regions giving a strong
entanglement between the two mechanical modes, even for large temperatures.
For instance, if the Bell measurement is performed after an interaction time
$t\sim 0.8\mu $s, one gets an optimal entanglement measure $E_{\mathcal{N}%
}^{out}\sim 0.88$, which does not depend on temperature (see the maximum in
Fig.~\ref{Plot}(a)). This is an effect of the quantum interference which
holds provided that the mechanical damping time $\Gamma ^{-1}$ is much
longer than the duration of the laser pulse and therefore the unitary
treatment of the process is valid. However, the \emph{living time} of the
generated entanglement will depend on the damping time $\Gamma ^{-1}$ scaled
by a factor depending on temperature. This decoherence time is given by $%
\gamma ^{-1}=\Gamma ^{-1}\ln [(2\bar{n}+1-2\eta _{out}^{-})(2\bar{n}%
+e^{-1}-2e^{-1}\eta _{out}^{-})^{-1}]$, which gives the limits $\gamma
^{-1}\simeq 1/\Gamma $ for $T\longrightarrow 0$ and $\gamma ^{-1}\simeq
\hbar \Omega (2\Gamma KT)^{-1}$ for $KT\gg \hbar \Omega $. Therefore, even
for long $\Gamma ^{-1}$, the entanglement living time will be affected by
temperature and it will be very short for high temperatures. For instance,
if $\Omega \simeq 5\times 10^{8}$ s$^{-1}$ and $\Gamma ^{-1}\simeq 1$s, one
has $\gamma ^{-1}\simeq 3$ $\mu $s at $T\simeq 300$ K, which has to be
compared with $\gamma ^{-1}\simeq 1$ ms at $T\simeq 1$ K. It is therefore
necessary to consider low temperatures in order to preserve the mechanical
entanglement, even if this assumption is not crucial for its generation.
Now, since the effective masses of the vibrational modes can be of the order
of $10^{-10}$ kg, this result proves the feasibility of entangling massive
macroscopic objects combining radiation pressure effects and simple optical
homodyne measurements. Actually, this performance can be further improved by
assisting the Bell measurement on the Stokes modes by additional heterodyne
measurements on the anti-Stokes modes (see Fig.~\ref{setup1color}). In fact,
heterodyne measurements are the Gaussian conditional measurements optimizing the entanglement
between the oscillator and the Stokes mode \cite{optim}. Adopting
this modified protocol, the output state $\rho _{out}$ will be again
Gaussian, but with a modified CM $V_{out}$, as given by Eq.~(\ref{Transf})
with the replacements: $a=\mathcal{B}+1/2-\mathcal{D}^{2}/(\mathcal{E}+1),$ $%
c=\mathcal{A}+1/2-\mathcal{F}^{2}/(\mathcal{E}+1)$ and $d=-d^{\prime }=%
\mathcal{C}+\mathcal{D}\mathcal{F}/(\mathcal{E}+1)$. The improvement is
shown in Fig.~\ref{Plot}(b) where the entanglement time regions are wider
and the maximum achievable entanglement is larger than before ($E_{\mathcal{N%
}}^{out}\sim 1.1$ at $t\sim 1\mu $s).

An important related issue is the experimental detection of the generated
macroscopic entanglement. In general, proving the existence of entanglement
between two oscillators requires the measurement of \emph{two} different
linear combinations of quadratures, like the relative distance $\hat{X}%
_{rel}\equiv \hat{x}_{a}-\hat{x}_{b}$ and the total momentum $\hat{P}%
_{tot}\equiv \hat{p}_{a}+\hat{p}_{b}$ \cite{CRITERIO,DuanPRL}. However, in
the present case, the final Gaussian state of Alice and Bob, both in the
assisted and non-assisted case, is a symmetrical channel where the variance
of relative distance and total momentum are \emph{equal}, i.e., $\langle
(\Delta \hat{X}_{rel})^{2}\rangle =\langle (\Delta \hat{P}_{tot})^{2}\rangle
$ (from Eq.~(\ref{Transf}) with $d=-d^{\prime }$), and, in particular, $%
\langle (\Delta \hat{X}_{rel})^{2}\rangle =2\eta _{out}^{-}$ (from Eq.~(\ref
{etam})). If the two micromechanical oscillators are highly reflecting
mirrors, they can be used as end mirrors of a Fabry-Perot cavity. It is
known that when this cavity is resonantly driven by an intense laser field,
the detection of the phase quadrature of the cavity output provides a
real-time quantum non-demolition measurement of the mirror relative distance
$\hat{X}_{rel}$ \cite{qnd}. Thus, if the cavity is driven soon after
Charlie's measurements and one performs a homodyne detection of the output
field, one can measure the variance $\langle (\Delta \hat{X}%
_{rel})^{2}\rangle $ and, therefore, the log-negativity of the swapped state
via Eq.~(\ref{logneg}). If the two oscillators are different, the
final state of Alice and Bob is no more symmetrical and one
has to measure both $\langle (\Delta \hat{X}_{rel})^{2}\rangle $ and $%
\langle (\Delta \hat{P}_{tot})^{2}\rangle $ in order to surely
detect entanglement (e.g., see Ref.~\cite{pinard}). Alternatively
one can measure $\hat{x}$ and $\hat{p}$ of each oscillator by
shining a second, intense ``reading'' laser pulse on it, and
exploiting again the same optomechanical scattering process. In
fact, as shown in Ref.~\cite{PRLePRA}, by performing a heterodyne
measurement of the linear combination of the two sidebands
$c-c'^{\dagger}$, one can reconstruct the oscillators' state.
Therefore one can detect the generated mechanical entanglement
provided that the living time of the entanglement, $\gamma^{-1}$,
is long enough and this is achievable at cryogenic temperatures.
In fact, as we have seen above, the thermal environment makes the
mechanical entanglement decay with a relaxation time which can be
quite long at low enough $T$ ($\gamma ^{-1}\simeq 1$ ms at
$T\simeq 1$ K if $\Omega \simeq 5\times 10^{8}$ s$^{-1}$ and
$\Gamma ^{-1}\simeq 1$s).

In conclusion, we have presented a scheme for entangling two
micromechanical oscillators by entanglement swapping. The scheme
exploits the optomechanical entanglement between each oscillator
and the reflected sideband modes of an intense laser field, which
can be generated by radiation pressure. Optical measurements are
then able to swap this entanglement to the oscillators, with the
non-trivial effect of changing the entanglement from
optomechanical to purely mechanical. To study the scheme, we have
first solved the general problem of entanglement swapping between
generic bipartite Gaussian states, providing simple input-output
formulas.

This work has been supported by MIUR (PRIN-2003): ``Schemes for
exploiting entanglement in optomechanical devices'' and by the
European Commission through FP6/2002/IST/FETPI SCALA: ``Scalable
Quantum computing with Light and Atoms'', Contract No. 015714.


\begin{thebibliography}{99}
\bibitem{Roukes03}  X. M. H. Huang \textit{et al.}, Nature (London) \textbf{%
421}, 496 (2003).

\bibitem{cleland03}  R. G. Knobel and A. N. Cleland, Nature (London) \textbf{%
424}, 291 (2003); M. D. LaHaye \textit{et al.}, Science \textbf{304}, 74
(2004).


\bibitem{Armour03}  A. D. Armour \textit{et al.}, Phys. Rev. Lett. \textbf{88%
}, 148301 (2002).

\bibitem{zoller}  L. Tian and P. Zoller, Phys. Rev. Lett. \textbf{93},
266403 (2004).


\bibitem{Marshall03}  W. Marshall \textit{et al.}, Phys. Rev. Lett. \textbf{%
91}, 130401 (2003);

\bibitem{eisert}  J. Eisert \textit{et al.}, Phys. Rev. Lett. \textbf{93},
190402 (2004).

\bibitem{CRITERIO}  S. Mancini \textit{et al.}, Phys. Rev. Lett. \textbf{88}%
, 120401 (2002).

\bibitem{Peng03}  J. Zhang \textit{et al.}, Phys. Rev. A \textbf{68}, 013808
(2003).

\bibitem{PRLePRA}  S. Mancini \textit{et al.}, Phys. Rev. Lett. \textbf{90},
137901 (2003); S. Pirandola \textit{et al.,} Phys. Rev. A \textbf{68},
062317 (2003).

\bibitem{Carmon05}  T. Carmon \textit{et al.}, Phys. Rev. Lett. \textbf{94},
223902 (2005).

\bibitem{vanloock}  P. Van Loock and S. L. Braunstein, Phys. Rev. A, \textbf{61}, 010302(R) (2000).


\bibitem{Salerno1}  G. Adesso \textit{et al.}, Phys. Rev. A \textbf{70},
022318 (2004).

\bibitem{DuanPRL}  L.-M. Duan \textit{et al.}, Phys. Rev. Lett. \textbf{84},
2722 (2000).



\bibitem{Cirac} G. Giedke \textit{et al.}, Phys. Rev. Lett. \textbf{91}, 107901
(2003); M. M. Wolf \textit{et al.}, Phys. Rev. A \textbf{69},
052320 (2004); M. M. Wolf \textit{et al.}, Phys. Rev. Lett.
\textbf{92}, 087903 (2004).


\bibitem{optim} S. Pirandola \textit{et al.}, Phys. Rev. A \textbf{71}, 042326 (2005).

\bibitem{qnd}  K. Jacobs \textit{et al.}, Phys. Rev. A \textbf{49}, 1961
(1994).

\bibitem{pinard}  M. Pinard \textit{et al.}, Europhys. Lett. \textbf{72}, 747 (2005).




\end{thebibliography}
\end{document}